\documentclass[aps,showpacs,preprintnumbers,eqsecnum,amsmath,amssymb]{revtex4}

\usepackage{graphics,epsfig}
\usepackage{graphicx}
\usepackage{dcolumn}
\usepackage{bm}

\newcommand{\be}{\begin{equation}}
\newcommand{\ee}{\end{equation}}
\newcommand{\ba}{\begin{eqnarray}}
\newcommand{\ea}{\end{eqnarray}}

\begin{document}
\title{Late-time behavior of massive Dirac fields in a Schwarzschild background}
\author{Jiliang Jing} \email{jljing@hunnu.edu.cn}
\affiliation{ Institute of Physics and  Department of Physics, \\
Hunan Normal University,\\ Changsha, Hunan 410081, P. R. China }

\vspace*{0.2cm}
\begin{abstract}
\vspace*{0.2cm}

The late-time tail behavior of massive Dirac fields is
investigated in the Schwarzschild black-hole geometry and the
result is compared with that of the massive scalar fields. It is
shown that in the intermediate late times there are three kinds of
differences between the massive Dirac and scalar fields, (I) the
asymptotic behavior of massive Dirac fields is dominated by a
decaying tail without any oscillation, but the massive scalar
field by a oscillatory inverse power-law  decaying tail, (II) the
dumping exponent for the massive Dirac field depends not only on
the multiple number of the wave mode but also on the mass of the
Dirac field, while that for the massive scalar field depends on
the multiple number only, and (III) the decay of the massive Dirac
field is slower than that of the massive scalar field.

\end{abstract}

\vspace*{0.4cm}

\maketitle

\section{INTRODUCTION}

The dynamical physical mechanism responsible for the relaxation of
perturbation fields outside a black hole and the decay rates of
the various perturbations have been extensively studied
\cite{3}-\cite{Ching} since Wheeler introduced the no-hair theorem
in the early 1970s \cite{1,2}. The massless neutral external
perturbations were first studied by Price, and it was found that
the late-time behavior for a fixed ~$r$ is dominated by the factor
~$t^{-(2l+3)}$ for each multiple moment ~$l$ \cite{3}. The
massless charged scalar field was studied in Refs.
\cite{4}-\cite{6} and the conclusion was that a charged hair decay
slower than a neutral one, i.e., the charged scalar hair outside a
charged black hole is dominated by a ~$t^{-(2l+2)}$ tail. The
massless late-time tail for the gravitational, electromagnetic,
neutrino and scalar perturbations had also been considered in the
case of the Kerr black holes in Refs. \cite{7}-\cite{9}. On the
other hand, many authors found that the analysis of massive fields
is also physically important since massive fields can cause
interesting phenomena which are qualitatively different from the
massless case. The evolution of a massive scalar field in the
Schwarzschild background was analyzed by Starobinskii and Novikov
\cite{Starobinskii}, and they found that, because of the mass
term, there are poles in the complex plane closer to the real axis
than in the massless case, which leads to inverse power-law
behavior with smaller indices than the massless case. Hod and
Piran \cite{13} pointed out that, if the field mass $\mu$ is
small, namely ~$\mu M\ll 1 $, the oscillatory inverse power-law
behavior
\begin{eqnarray}
\label{1}&& \Phi\sim t^{-(l+3/2)}\sin(\mu t),
\end{eqnarray}
dominates as the intermediate late-time tails in the
Reissner-Nordstr\"{o}m background. We \cite{Jing1} recently
investigated the late-time tails of the massless and the
self-interacting (massive) scalar fields in a stationary
axisymmetric Einstein-Maxwell-dilaton-axion black-hole geometry,
and found that the dumping exponents is independent of the
rotation parameter and the dilaton of the black hole.

Although much attention has been paid to the study of the
late-time behaviors of the scalar, gravitational, electromagnetic
in static and stationary black-hole backgrounds, however, to my
best knowledge, at the moment the late-time evolution of the Dirac
fields has not been investigated. The aim of this paper is to
study the intermediate late-time tail behavior of the massive
Dirac fields in the Schwarzschild black-hole background and to see
whether or not special properties exist in this case.

The plan of the paper is  as follows. In Sec.2 the decoupled
massive Dirac equations in the  Schwarzschild spacetime are
presented. In Sec.3  the black-hole Green's function  is
introduced by using the spectral decomposition method \cite{17}.
In Sec.4  the intermediate late-time evolution of the Dirac
massive fields in the Schwarzschild background is investigated.
The section V is devoted to a summary and conclusion. In the
appendix we study whether the conclusions might change if the
tortoise coordinate is defined in the conventional way.

\section{Dirac equation in the Schwarzschild spacetime}
 \vspace*{0.2cm}

Dirac equation in a general background spacetime can be expressed
as \cite{Brill}
 \begin{eqnarray}
[\gamma^a e_a^\mu(\partial _\mu+\Gamma_\mu)+\mu]\Psi=0,
\label{Di}
\end{eqnarray}
where $\mu$ is the mass of the Dirac field, $\gamma^a$ is the
Dirac matrix, $e_a^\mu$ is the inverse of the tetrad $e_\mu^a$,
and $\Gamma_\mu $ is the spin connection, which is defined as
$\Gamma_\mu= \frac{1}{8}[\gamma^a,\gamma^b] e_a^\nu e_{b\nu;\mu}$.
For the Schwarzschild black hole
\begin{eqnarray}    ds^2=-f dt^2+\frac{1}{f}dr^2+r^2(d
\theta^2 +sin^2\theta d\varphi^2),
\end{eqnarray}
with
\begin{eqnarray}
    f=1-\frac{2M}{r},
\end{eqnarray}
where the parameters $M$ represents the mass of the black hole, we
can take the tetrad as
\begin{eqnarray}
    e_\mu^a=diag(\sqrt{f}, \frac{1}{\sqrt{f}}, r, r \sin \theta).
\end{eqnarray}

Introducing an ansatz
\begin{eqnarray}
    \Psi=f^{-\frac{1}{4}} \left(
\begin{array}{c}
\frac{i F_1^{(\pm)}(r,t)}{r}\phi^{\pm}_{jm}(\theta, \varphi) \\
\frac{F_2^{(\pm)}(r,t)}{r}\phi^{\mp}_{jm}(\theta, \varphi)
\end{array}\right),
\end{eqnarray}
with
\begin{eqnarray}
    \phi^{+}_{jm}=\left(
\begin{array}{c}
\sqrt{\frac{j+m}{2 j}}Y^{m-1/2}_l \\
\sqrt{\frac{j-m}{2 j}}Y^{m+1/2}_l
\end{array}\right) \ \ \ \ \ \ \ \ \ \ \ \  for \ \ j=l+\frac{1}{2},
\nonumber
\end{eqnarray}
\begin{eqnarray}
    \phi^{-}_{jm}=\left(
\begin{array}{c}
\sqrt{\frac{j+1-m}{2 j+2}}Y^{m-1/2}_l \\
-\sqrt{\frac{j+1+m}{2 j+2}}Y^{m+1/2}_l
\end{array}\right) \ \ \ \ \ \ for \ \ j=l-\frac{1}{2}, \nonumber
\end{eqnarray}
Cho \cite{Cho} find that the cases for $(+)$ and $(-)$ in the
functions $F_1^{\pm}$ and $F_2^{\pm}$ can be put together, and
then the decoupled equations can be written as
\begin{eqnarray}
    \left(\frac{\partial^2 }{\partial \hat{r}_*^2}-\frac{\partial^2}
    {\partial t^2}-V_i\right)F_i(\hat{r}_*,t)&=&0, ~~~~~~~~~~
    \text{i=1,2}
    \label{even}
\end{eqnarray}
where
\begin{eqnarray}\label{tor}
&&\hat{r}_*=r+2M \ln \left(\frac{r}{2M}-1\right)+\frac{1}{2
\omega} tan^{-1}\left(\frac{\mu ~r}{|k|
}\right)\\
&&V_{1,2}=\pm \frac{d W}{d\hat{r}_*}+W^2, \label{V1}
\end{eqnarray}
with
\begin{eqnarray}
W&=&\frac{\Delta^{1/2}(k^2+\mu^2r^2)^{3/2}}{r^2(k^2+\mu^2r^2)+\mu
k \Delta/2\omega}, \label{V2}
\end{eqnarray}
where $\Delta=r(r-2M)$. Here $k$ goes over all positive and
negative integers. Positive integers represent the $(+)$ case of
Eq. (\ref{V1})  with $k=j+1/2$ and $j=l+1/2$, while negative
integers represent the $(-)$ case of Eq. (\ref{V1}) with
$k=-(j+1/2)$ and $j=l-1/2$. Equation (\ref{V1}) shows that the
potentials $V_1$ and $V_2$ are related to the metric function
$\sqrt{f}$ which differs from potentials for the scalar,
electromagnetic and gravitational fields \cite{Jing2,Jing3}.  We
will use the potential (\ref{V1}) to study the late-time tail
behavior of the massive Dirac fields in the Schwarzschild
black-hole geometry.

It is easy to see that near the event horizon the effective
potential $V_{1,2}$ becomes zero. Then the radial asymptotic
solutions can be expressed as
\begin{eqnarray}
R_i\simeq e^{\pm i \omega \hat{r}_*}=e^{\pm i\omega r_*}e^{\pm
\frac{i}{2}tan^{-1}\left(\frac{\mu r}{|k|}\right)}, \label{Solu1}
\end{eqnarray}
where $r_*=\int 1/f dr$ is the usual tortoise coordinate. By
comparing Eq. (\ref{Solu1}) with the scalar field solutions
$e^{\pm i \omega r_*}$ \cite{13} we know that the Dirac field
possesses a additional factor $e^{\pm
\frac{i}{2}tan^{-1}\left(\frac{\mu r}{|k|}\right)}$. The
appearance of the factor seems due to we used the ``tortoise"
coordinate (\ref{tor}). However, the factor still exists even if
we use the usual tortoise coordinate (see appendix for detail).

\section{The black hole Green's function}

It is well known that the time evolution of a wave field
$\Psi(r_*,t)$ follows from
\begin{eqnarray}
\label{9}&&\Psi(r_*,t)=\int [G(r_*,r_*';t)\partial_t
\Psi(r_*',0)+\partial_t G(r_*,r_*';t)\Psi(r_*',0)] dr_*',
\end{eqnarray}
where the black-hole (retarded) Green's function $G(r_*,r_*';t)$
is defined by
\begin{eqnarray}
\label{10}&&\left[\frac{\partial^{2}} {\partial
t^{2}}-\frac{\partial^{2}}{\partial r_*^{2}}+V(r)
\right]G(r_*,r_*';t)
 =\delta(t)\delta(r_*-r_*').
\end{eqnarray}
The causality condition gives us the initial condition
$G(r_*,r_*';t)=0$ for $t\leq0$. In order to get $G(r_*,r_*';t)$ we
use the Fourier transform
\begin{eqnarray}
\label{11}&&\tilde{G}(r_*,r_*';\omega) = \int_{0^{-}}^\infty
G(r_*,r_*';t)
 e^{i\omega t} dt.
\end{eqnarray}
The  Fourier transform  is analytic in the upper half
~$\omega-$plane, and the corresponding inversion formula is given
by
\begin{eqnarray}
\label{12}\ \ \ \ G(r_*,r_*';t)=\frac{1}{2\pi}
\int_{-\infty+ic}^{\infty+ic}\tilde{G}(r_*,r_*';\omega) e^{-i
\omega t}d\omega,
\end{eqnarray}
where $c$ is some positive constant.

We define two auxiliary functions $\tilde{\Psi}_{1} (r_*,\omega)$
and $\tilde{\Psi}_{2}(r_*,\omega)$ which are (linearly
independent) solutions to the homogeneous equation
\begin{eqnarray}
\label{16}&&\left[\frac{d^{2}}{dr_*^{2}}+\omega^{2}-V(r)
\right]\tilde{\Psi}_{i}(r_*,\omega)=0,\ \ i =1,2.
\end{eqnarray}
Let the Wronskian be
\begin{eqnarray}
\label{17}\ \ \ \ W(\omega)=W(\tilde{\Psi}_{1}, \tilde{\Psi}_{2})
=\tilde{\Psi}_{1}\tilde{\Psi}_{2,x}-\tilde{\Psi}_{2}
\tilde{\Psi}_{1,x},
\end{eqnarray}
and using the two solutions $\tilde{\Psi}_{1}$ and $
\tilde{\Psi}_{2}$, the black-hole Green's function can be
constructed as
\begin{eqnarray}
\label{18}\ \ \ \ \tilde{G}_{l}(r_*,r_*';\omega)=-\frac{1}
{W(\omega)} \left\{ \begin{array}{l}
\tilde{\Psi}_{1}(r_*,\omega)\tilde{\Psi}_{2}(r_*',
\omega),\ \ r_*<r_*'; \\
\tilde{\Psi}_{1}(r_*',\omega)\tilde{\Psi}_{2}(r_*,\omega), \ \
r_*>r_*'.
\end{array} \right.
\end{eqnarray}

\section{Late-time tails of the massive Dirac fields}

It is well known that massive tails exist even in a flat spacetime
\cite{19}. This phenomenon is related to the fact that different
frequencies forming a massive wave packet have different phase
velocities. One will see that at intermediate times the
backscattering from asymptotically far regions (which dominates
the tails of massless fields) is negligible compared to the flat
spacetime massive tails that appear here. Hod, Piran, and Leaver
\cite{13, 17} argued that the asymptotic massive tail is
associated with the existence of a branch cut (in
$\tilde{\Psi}_{2}$) placed along the interval $-\mu\leq
\omega\leq\mu$. This tail arises from the integral of the Green
function $\tilde{G}(r_*, r_*';\omega)$ around the branch (denoted
by $G^C(r_*, r_*';\omega)$). So our goal is to evaluate $G^C(r_*,
r_*';\omega)$.

Now we assume that both the observer and the initial data are
situated far away from the black hole so that $r\gg M$. One may
expand the wave-equation (\ref{16}) with the potential $V_{1}$
which is given by (\ref{V1}) (or $V_2$) for the massive Dirac
field as a power series in $M/r$ (neglecting terms of order
$O\left(\left(M/r\right)^2\right)$) as follows
\begin{eqnarray}
\label{m19}&&\left(\frac{d^{2}}{dr^{2}}+\omega^{2}-\mu^2+
\frac{4M\omega^{2}-2M\mu^2}{r}-\frac{k^2-k\omega
/\mu}{r^{2}}\right) \xi(r,\omega)=0,
\end{eqnarray}
with
\begin{eqnarray}
 \xi=\left(1-2
M/r\right)^{1/2}\left[1+\left(1-\frac{2M}{r}\right)\frac{\mu
|k|}{2\omega (k^2+\mu^2 r^2)}\right]^{-1/2} \tilde{\Psi}.
\end{eqnarray}
At great distances, the only difference between effective
potentials of the scalar and Dirac fields is that the term
$l(l+1)/r^2$ for the massive scalar field is now replaced by
$(k^2-k\omega/\mu)/r^2$ for the massive Dirac field. As we will
see later, it is the difference that leads to new result.

Let us introduce
\begin{eqnarray}
\label{m19a}&&  z=2 \sqrt{\mu^2- \omega^2} r=2\varpi r \\ &&
\xi=e^{-z/2}z^{1/2+b}\Phi, \\ &&
b^2=\frac{1}{4}+k^2-\frac{k\omega}{\mu}, \\ &&
a=\frac{M\mu^2}{\varpi}-2M\varpi.
\end{eqnarray}
Then Eq. (\ref{m19}) becomes the confluent hypergeometric equation
\begin{eqnarray}
\label{m19b}&& z\frac{d^2 \Phi}{dz^2}+(1+2
b-z)\frac{d\Phi}{dz}-(\frac{1}{2}+b-a)\Phi=0.
\end{eqnarray}
It follows that two basic solutions required in order to build the
Green's function can be expressed as
\begin{eqnarray}
\label{mgreen}&&\tilde{\psi}_1=Ae^{-\varpi r}(2\varpi r)
^{1/2+b}M(1/2+b-a, 1+2 b, 2\varpi r), \\
&&\tilde{\psi}_2=Be^{-\varpi r}(2\varpi r) ^{1/2+b}U(1/2+b-a, 1+2
b, 2\varpi r),
\end{eqnarray}
where $A$ and $B$ are normalization constants. The functions
$M(\tilde{a},\tilde{b},z)$ and $U(\tilde{a},\tilde{b},z)$
represent the two standard solutions to the confluent
hypergeometric equation \cite{Abram}. $U(\tilde{a},\tilde{b},z)$
is a many-valued function, i.e., there is a cut in
$\tilde{\psi}_2$.

\begin{figure}
\includegraphics[scale=0.6]{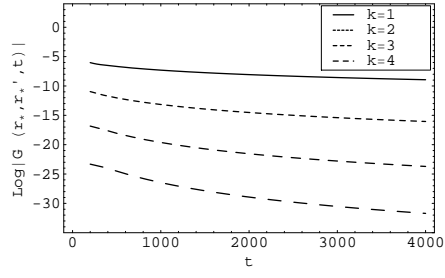}
\caption{\label{fig1}The figure describes $\ln|G^C(r_*,r_*',t)|$
versus $t$ for different $k$ with the same mass ($\mu=0.01$),
which shows that the dumping exponent depends on the multiple
number of the wave mode.}
\end{figure}

\begin{figure}
\includegraphics[scale=0.6]{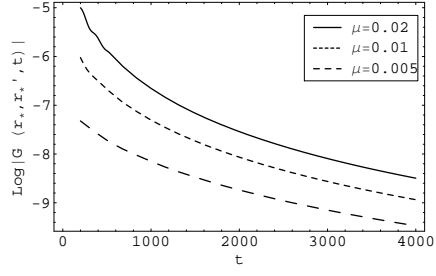}
\caption{\label{fig2}The  figure describes $\ln|G^C(r_*,r_*',t)|$
versus $t$ for different mass $\mu$ with the same multiple number
($k=1$), which indicates that the dumping exponent depends on the
mass $\mu$ of the Dirac fields.}
\end{figure}

\begin{figure}
\includegraphics[scale=0.55]{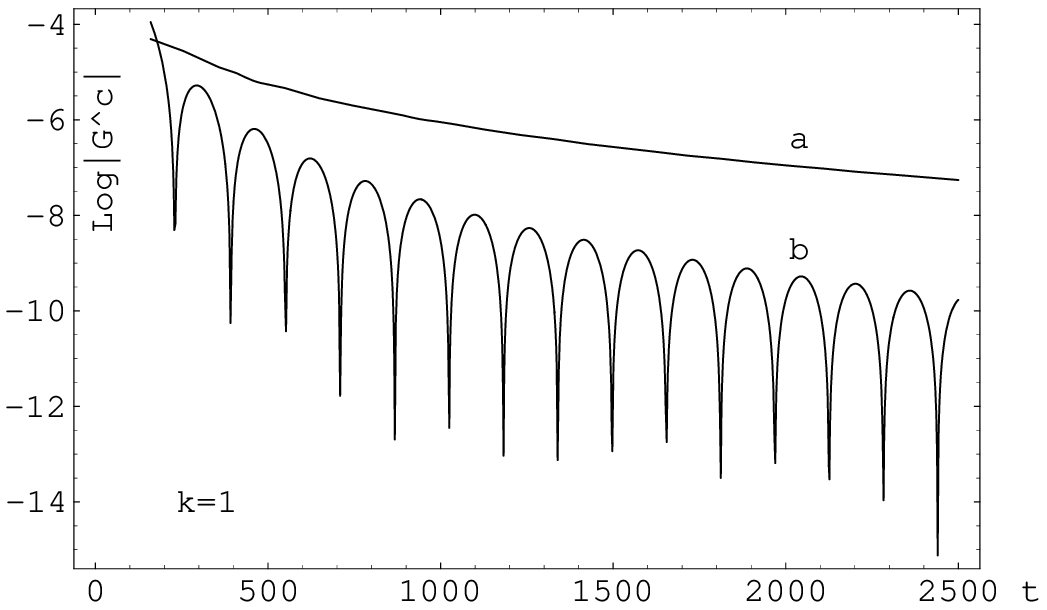}\hspace{0.5cm}%
\includegraphics[scale=0.55]{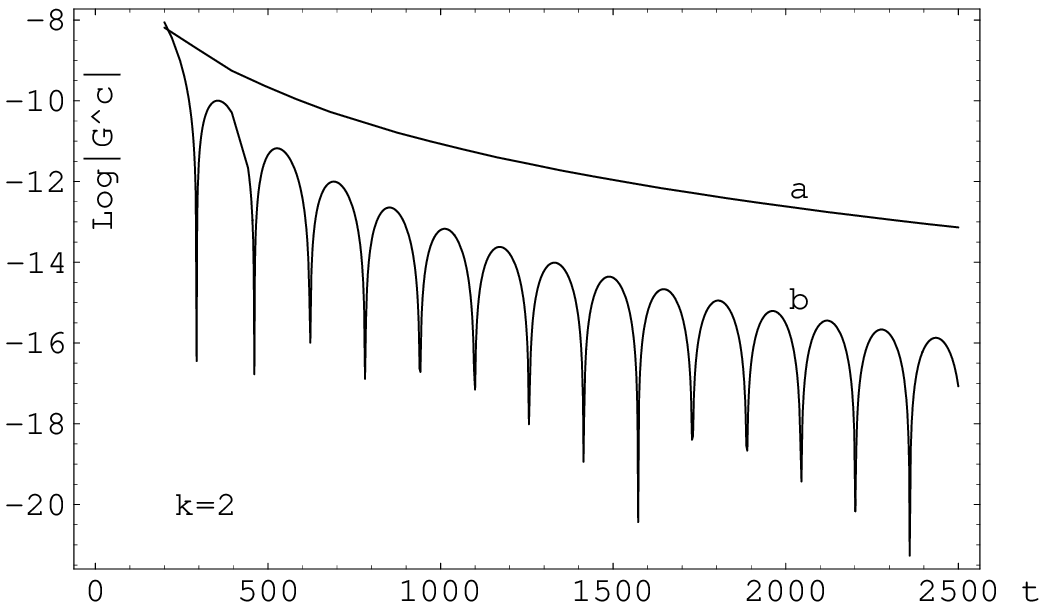}\\\vspace{0.2cm}
\includegraphics[scale=0.55]{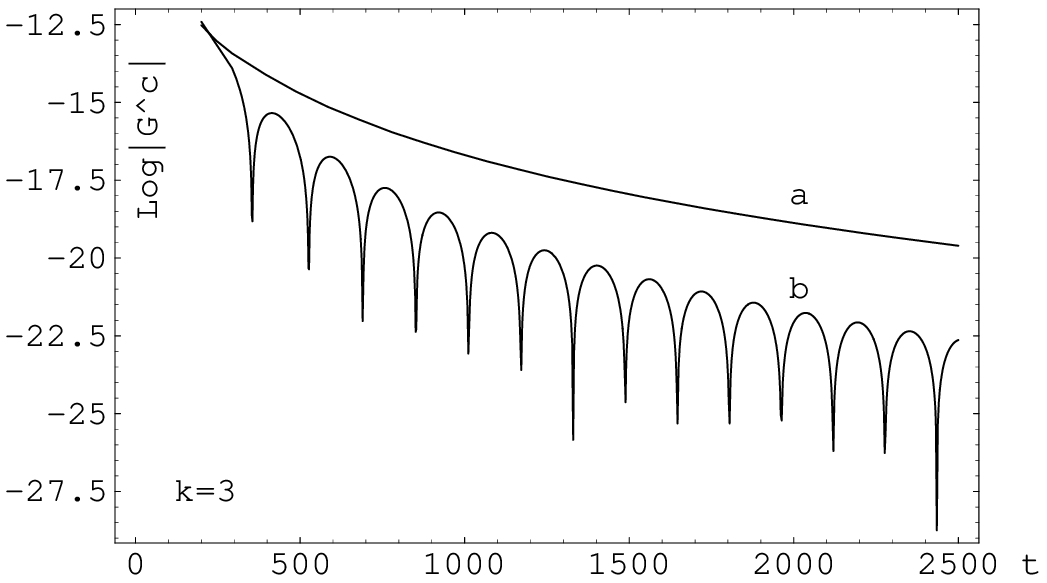}\hspace{0.5cm}%
\includegraphics[scale=0.55]{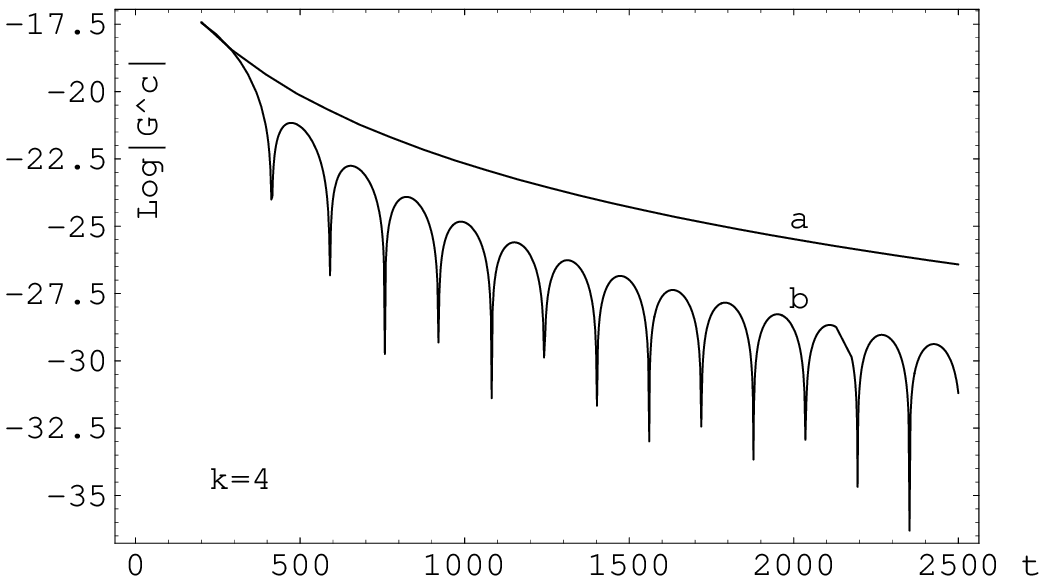}\\ \vspace{1.5cm}
\includegraphics[scale=0.55]{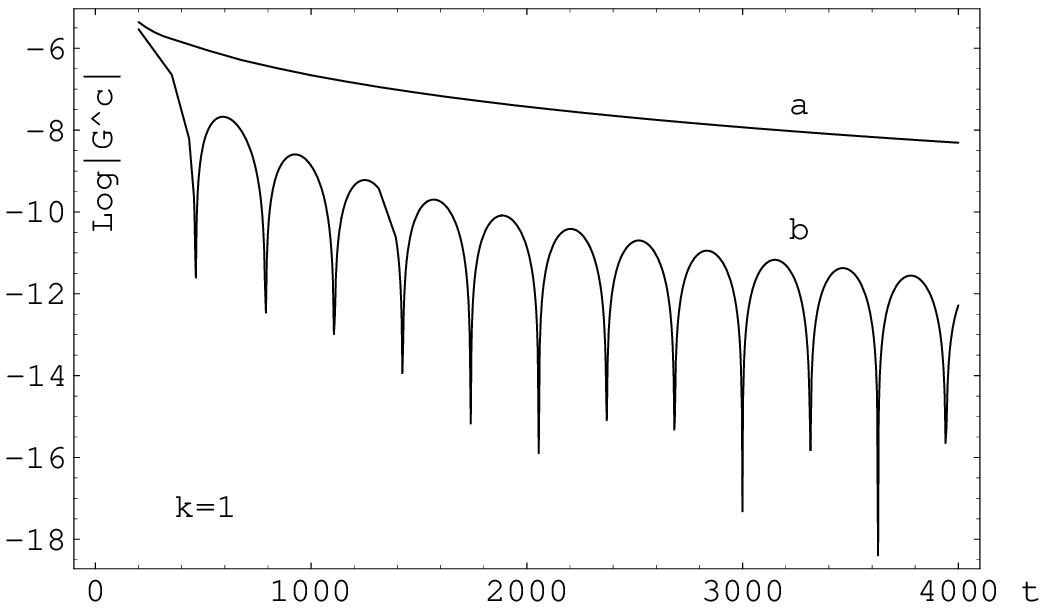}\hspace{0.5cm}%
\includegraphics[scale=0.55]{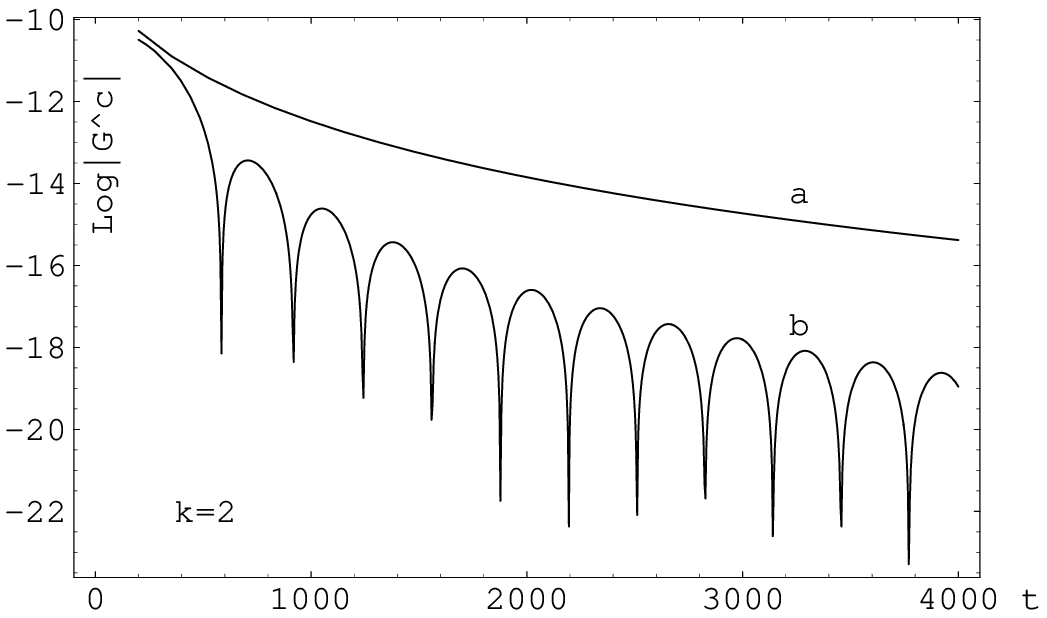}\\ \vspace{0.2cm}
\includegraphics[scale=0.55]{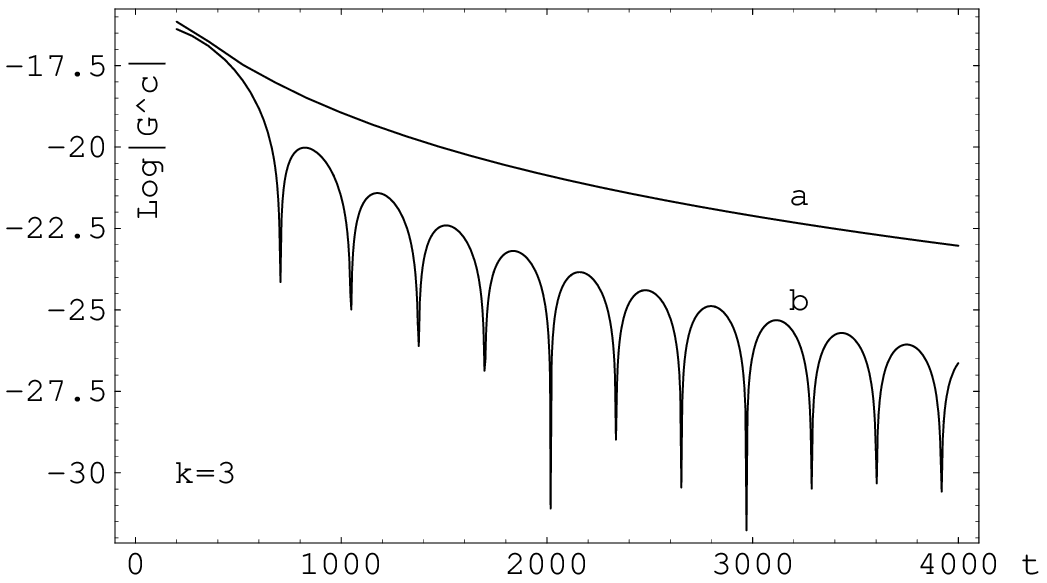}\hspace{0.5cm}%
\includegraphics[scale=0.55]{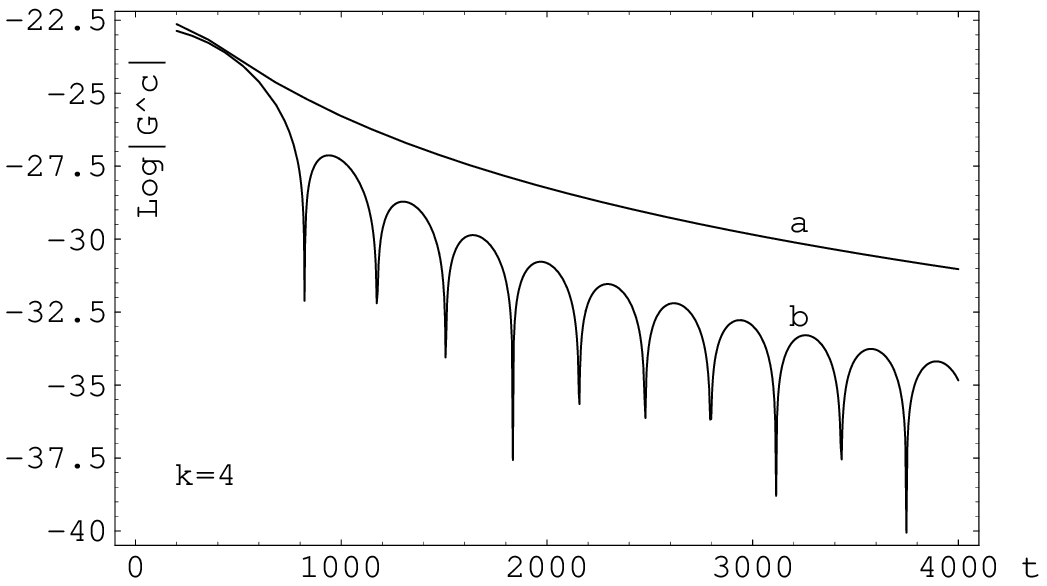}
\caption{\label{fig3} The above four figures describe
$\ln|G^C(r_*,r_*',t)|$ versus $t$ for $\mu=0.02$, and below four
for $\mu=0.01$. The dashing lines represent the result of the
Green function of the massive Dirac field with different $k$. For
comparing, we also show the corresponding result of the Scalar
field with solid lines. It is shown that the late-time behavior of
massive Dirac fields is dominated by a decaying tail without any
oscillation, and the decay of the massive Dirac field is slower
than that of the massive scalar field.}
\end{figure}

Using Eq. (\ref{12}), one finds that the branch cut contribution
to the Green's function is described by
\begin{eqnarray}
\label{mgreen1}G^C(r_*,r_*',t) &=& \frac{1}{2\pi}\int_{-\mu}^\mu
\left[\frac{\tilde{\psi}_1(r_*',\omega
e^{i\pi})\tilde{\psi}_2(r_*,\omega e^{i \pi})} {W(\omega e^{i
\pi})}-\frac{\tilde{\psi}_1(r_*',\omega)
\tilde{\psi}_2(r_*,\omega) }
{W(\omega )}\right] e^{-i\omega t} d\omega\nonumber \\
&=&\frac{1}{2\pi}\int_{-\mu}^\mu F(\varpi) e^{-i\omega t}d \omega.
\end{eqnarray}
Using the following relations
\begin{eqnarray}
\label{mrel1}  \tilde{\psi}_1(2\varpi r e^{i\pi}
)&=&Ae^{(1/2+b)i\pi}e^{-\varpi r}(2\varpi r)
^{1/2+b}M(1/2+b+a, 1+2 b, 2\varpi r),\nonumber \\
\tilde{\psi}_2(2\varpi r )&=& B\frac{\Gamma(-2
b)}{\Gamma(1/2-b-a)}e^{-\varpi r}(2\varpi r)^{1/2+b}M(1/2+b-a,
1+2b, 2\varpi r)\nonumber\\ &+&B\frac{\Gamma(2
b)}{\Gamma(1/2+b-a)}e^{-\varpi r}(2\varpi r)^{1/2-b}M(1/2-b-a,
1-2b, 2\varpi r),
\end{eqnarray}
we find
\begin{equation}
W(\varpi e^{i\pi})=-W(\varpi)=AB\frac{\Gamma(2b)}{\Gamma(1/2+b-a)}
4 b\varpi.
\end{equation}
and
\begin{eqnarray}
\label{F1}F(\varpi)&=&\frac{( r_*)^{\frac{1}{2}-b}(
r_*')^{\frac{1}{2}+b}e^{-\varpi (r_*+r_*')}}{2
b}\left[M(\frac{1}{2}+b+a,1+2b,2\varpi
r_*')M(\frac{1}{2}-b+a,1-2b, 2\varpi r_*) \right.  \nonumber
\\ &-& \left.  M(\frac{1}{2}+b-a,1+2b,2\varpi r_*')M(\frac{1}{2}-b-a,1-2b,2\varpi
r_*)\right] -\frac{\Gamma(-2b)\Gamma(\frac{1}{2}+b-a)}{\Gamma(2b)
\Gamma(\frac{1}{2}-b-a)}\nonumber
\\ &\times&\frac{(4\varpi^2 r_* r_*')
^{\frac{1}{2}+b}e^{-\varpi
(r_*+r_*')}}{4\varpi b}\left[M(\frac{1}{2}+b-a,1+2b,2\varpi
r_*')M(\frac{1}{2}+b-a,1+2b, 2\varpi r_*) \right.  \nonumber
\\ &+& \left.  e^{(1+2b)i\pi}M(\frac{1}{2}+b+a,1+2b,2\varpi
r_*')M(\frac{1}{2}+b+a,1+2b, 2\varpi r_*) \right].
\end{eqnarray}
If we further assume that both the observer and the initial data
are situated in the region $ M\ll r \ll M/(\mu M)^2$ and we are
interested in the intermediate asymptotic behavior of the Dirac
field $[M\ll r\ll t \ll M/(M\mu)^2]$, we find that the frequency
range $\varpi=O(\sqrt{\mu/t})$, which gives the dominant
contribution to the integral, implies $a\ll 1$. Eq. (\ref{m19})
shows that $a$ originates from the $1/r$ term which describes the
effect of backscattering off the spacetime curvature. Thus, the
backscattering off the curvature from the asymptotically regions
is negligible for the case $a\ll 1$. So, we get
\begin{eqnarray}
\label{F2}F(\varpi)&\approx&
-\frac{\Gamma(-2b)\Gamma(\frac{1}{2}+b)}{\Gamma(2b)\Gamma(\frac{1}{2}-b)}
\frac{(2\varpi r_*)^{1/2+b}(2\varpi r_*')^{1/2+b}e^{-\varpi
(r_*+r_*')}}{4\varpi b}\nonumber
\\ && (1+e^{(1+2b)i\pi})M(\frac{1}{2}+b,1+2b,2\varpi r_*')M(\frac{1}{2}+b,1+2b,
2\varpi r_*).
\end{eqnarray}
Noticing that $M(\tilde{a},\tilde{b},z)\approx 1$ as $z\rightarrow
0$, we have
\begin{eqnarray}
\label{F3}F(\varpi)&\approx& -\frac{1+e^{(1+2b)i\pi}}{4\varpi
b}\frac{\Gamma(-2b)\Gamma(\frac{1}{2}+b)}{\Gamma(2b)\Gamma(\frac{1}{2}-b)}
(2\varpi)^{1+2b}(r_*r_*')^{1/2+b}\nonumber \\ &&
=\frac{\pi}{sin(\pi b)}\frac{1+e^{(1+2b)i
\pi}}{2^{1+2b}b^2}\frac{\varpi^{2b}}{\Gamma(b)^2}(r_*r_*')^{1/2+b},
\end{eqnarray}
where we used the formulae
$\Gamma(2z)=\frac{1}{\sqrt{2\pi}}2^{2z-1/2}\Gamma(z)\Gamma(1/2+z)$
and $\Gamma(-z)=-\frac{\pi}{sin(\pi z)}\frac{1}{z\Gamma(z)}$.
Substituting Eq. (\ref{F3}) into the Eq. (\ref{mgreen1}), we
obtain
\begin{eqnarray}
\label{mgreen2}G^C(r_*,r_*',t) &=& \frac{1}{4}\int_{-\mu}^\mu
\frac{1}{sin(\pi b)}\frac{1+e^{(1+2b)i \pi}}{2^{2b}b^2}\frac{
(r_*r_*')^{1/2+b}}{\Gamma(b)^2}(\mu^2-\omega^2)^{b}e^{-i\omega t}d
\omega.
\end{eqnarray}
Unfortunately, the integral can not be evaluated analytically
since the parameter $b=\sqrt{1/4+k^2-k\omega/\mu}$ depends on
$\omega$. However, we can work out the integral numerically and
the results are presented in figures \ref{fig1}-\ref{fig3} with
the form $\ln|G^C(r_*,r_*',t)|$ versus $t$. Figure \ref{fig1}
describes different $k$ with the same mass ($\mu=0.01$), which
shows that the dumping exponent depends on the multiple number of
the wave mode.  Figure \ref{fig2} gives different mass $\mu$ with
the same multiple number ($k=1$), which indicates that the dumping
exponent depends on the mass $\mu$ of the Dirac fields. In Fig.
\ref{fig3} the lines (a) represent the result of the Green
function (\ref{mgreen2}) of the massive Dirac field with different
$k$, and lines (b) show the corresponding result of the massive
scalar field.  We learn from Fig. \ref{fig3} that late-time
behavior of massive Dirac fields is dominated by a decaying tail
without any oscillation, and the decay of the massive Dirac field
is slower than that of the massive scalar field.

\section{Summary and discussions}

We have studied the intermediate late-time tail behavior of
massive Dirac fields in the Schwarzschild black-hole geometry  and
the result is compared with that of the massive scalar fields. We
know from the figures that there are three differences between the
massive Dirac and scalar fields due to the fact that the parameter
$b=\sqrt{1/4+k^2-k \omega/\mu }$ is related to $\omega$ for the
massive Dirac fields while $b=l+1/2$ for the massive scalar
fields, (I) the late-time behavior of massive Dirac fields is
dominated by an inverse power-law decaying tail without any
oscillation, but the massive scalar field by an oscillatory
decaying tail, (II) the dumping exponent (i.e., $\alpha$ in term
$1/t^\alpha$) for the massive Dirac field depends not only on the
multiple number of the wave mode but also on the mass of the Dirac
field, while $\alpha=(l+3/2)$ for the massive scalar field depends
on the multiple number only,  and (III) the decay of the massive
Dirac field is slower than that of the massive scalar field.

\begin{acknowledgments}This work was supported in part by the
National Natural Science Foundation of China under Grant No.
10275024; the FANEDD under Grant No. 200317; the Hunan Provincial
Natural Science Foundation of China under Grant No. 04JJ3019; and
the National Basic Research program of China under grant No.
2003CB716300.
\end{acknowledgments}

\appendix
\section{may the conclusions change in the usual tortoise
coordinates?}

The ``tortoise" coordinate (\ref{tor}) is a function of the
background geometry and the test field, which is different from
the usual tortoise one, $r_*=\int \frac{d r}{f}$.  Might the
conclusions of this paper change if we use usual tortoise
coordinate instead of the coordinate (\ref{tor})? In the appendix
we will address this question.

The Dirac equations are given by \cite{Page}
\begin{eqnarray}
   &&\sqrt{2}\nabla_{BB'}P^B+i\mu \bar{Q}_{B'}=0, \nonumber \\
   &&\sqrt{2}\nabla_{BB'}Q^B+i\mu \bar{P}_{B'}=0,
\end{eqnarray}
where $\nabla_{BB'}$ is covariant differentiation, $A_{BB'}$ is
the electromagnetic field potential, $P^B$ and $Q^B$ are the
two-component spinors, and $\mu $ is the particle mass. In the
Newman-Penrose formalism \cite{Newman} the equations become
\begin{eqnarray}\label{np}
   &&(D+\epsilon-\rho )P^0+
   (\bar{\delta}+\pi-\alpha )P^1=2^{-1/2}i\mu  \bar{Q}^{1'},\nonumber \\
   &&(\triangle+\mu -\gamma )P^1+
   (\delta+\beta -\tau )P^0=-2^{-1/2}i\mu
    \bar{Q}^{0'}, \nonumber\\
   &&(D+\bar{\epsilon}-\bar{\rho} )\bar{Q}^{0'}+
   (\delta+\bar{\pi}-\bar{\alpha} )\bar{Q}^{1'}=-2^{-1/2}i\mu  P^{1},\nonumber \\
  &&(\triangle+\bar{\mu} -\bar{\gamma} )\bar{Q}^{1'}+
   (\bar{\delta}+\bar{\beta} -\bar{\tau} )\bar{Q}^{0'}=2^{-1/2}i\mu  P^{0},
\end{eqnarray}
For the Schwarzschild spacetime the null tetrad can be taken as
\begin{eqnarray}
  &&l^\mu=(\frac{r^2}{\Delta}, ~0, ~0, ~0 ), \nonumber \\
  &&n^\mu=\frac{1}{2}(1, ~-\frac{\Delta}{r^2}, ~0, ~0)\nonumber \\
  &&m^\mu=\frac{1}{\sqrt{2} r}\left(0, ~0, ~1, \frac{i}{sin\theta}\right),
\end{eqnarray}
Then, if we set
\begin{eqnarray}
&&P^0=\frac{1}{r}f_1(r,\theta)e^{-i(\omega t-m\varphi)}, \nonumber \\
&&P^1=f_2(r,\theta)e^{-i(\omega t-m\varphi)}, \nonumber \\
&&\bar{Q}^{1'}=g_1(r,\theta)e^{-i(\omega t-m\varphi)}, \nonumber \\
&&\bar{Q}^{0'}=-\frac{1}{r}g_2(r,\theta)e^{-i(\omega t-m\varphi)},
\end{eqnarray}
where $\omega$ and $m$ are the energy and angular momentum of the
Dirac particles, Eq. (\ref{np}) can be simplified as
\begin{eqnarray} \label{DD}
&&{\mathcal{D}}_0 f_1+\frac{1}{\sqrt{2}}{\mathcal{L}}_{1/2}
f_2=\frac{1}{\sqrt{2}}i\mu r g_1,
\nonumber \\
&&{\Delta \mathcal{D}}_{1/2}^{\dag}
f_2-\sqrt{2}{\mathcal{L}}_{1/2}^{\dag} f_1=-\sqrt{2}i\mu r g_1,
\nonumber \\
&&{\mathcal{D}}_0 g_2-\frac{1}{\sqrt{2}}{\mathcal{L}}_{1/2}^{\dag}
g_1=\frac{1}{\sqrt{2}}i\mu r f_2,
\nonumber \\
&& {\Delta \mathcal{D}}_{1/2}^{\dag}
g_1+\sqrt{2}{\mathcal{L}}_{1/2} g_2=-\sqrt{2}i\mu r f_1,
\end{eqnarray}
with
 \begin{eqnarray}
 &&{\mathcal{D}}_n=\frac{\partial}{\partial r}-\frac{i K}
 {\bigtriangleup}+2n\frac{r-M}{\bigtriangleup},\nonumber \\
 &&{\mathcal{D}}^{\dag}_n=\frac{\partial}{\partial r}+\frac{i K}
 {\bigtriangleup}+2n\frac{r-M}{\bigtriangleup},\nonumber \\
 &&{\mathcal{L}}_n=\frac{\partial}{\partial \theta}+\frac{m}{\sin \theta }
 +n\cot \theta,\nonumber \\
 &&{\mathcal{L}}^{\dag}_n=\frac{\partial}{\partial \theta}-\frac{m}{\sin \theta }
 +n\cot \theta, \nonumber \\
 &&K=r^2\omega.\label{ld}
 \end{eqnarray}
It is now apparent that the variables can be separated by the
substitutions
 \begin{eqnarray}
f_1=R_{-1/2}(r)S_{-1/2}(\theta),\nonumber \\
f_2=R_{+1/2}(r)S_{+1/2}(\theta),\nonumber \\
g_1=R_{+1/2}(r)S_{-1/2}(\theta),\nonumber \\
g_2=R_{-1/2}(r)S_{+1/2}(\theta).
 \end{eqnarray}
Thus, we have\cite{Chand}
 \begin{eqnarray} \label{DDD}
\left[\Delta
{\mathcal{D}}^{\dag}_{1/2}{\mathcal{D}}_0-\frac{i\mu\Delta}{\lambda+i\mu
r}{\mathcal{D}} _0-(\lambda^2+\mu^2 r^2)\right]R_{-1/2}=0,
 \end{eqnarray}
and $\sqrt{\Delta}R_{+1/2}$ satisfies the complex-conjugate
equation. The decoupled equations can then be explicitly expressed
as
\begin{eqnarray} \label{T1}
&& \sqrt{\Delta} \frac{d }{d r}\left(\sqrt{\Delta}
\frac{dR_{-1/2}}{d r}\right)-\frac{i\mu\Delta}{\lambda+i\mu
r}\frac{d R_{-1/2}}{d r}+P_- R_{-1/2}
 =0, \\  \label{T2}
&& \frac{1}{\sqrt{\Delta}} \frac{d }{d r}\left(\Delta^{3/2}
\frac{dR_{+1/2}}{d r}\right)+\frac{i\mu\Delta}{\lambda-i\mu
r}\frac{d R_{+1/2}}{d r}+P_+ R_{+1/2}
 =0,
 \end{eqnarray}
with
 \begin{eqnarray}
&& P_-=\frac{K^2+i   (r-M) K}{ \Delta}
 -2 i  \omega  r
 -\frac{\mu K}{\lambda+i\mu r}
 -\mu^2r^2-\lambda^2,\\
 &&P_+=\frac{K^2-i (r-M) K}{ \Delta} +2 i  \omega  r+2 s
 +\frac{\mu(i(r-M)-K)}{\lambda-i\mu r}
 -\mu^2r^2 -\lambda^2,
 \end{eqnarray}
where $\lambda^2=(l-s)(l+s+1)$ is the separation constant.
Introducing an usual tortoise coordinate
\begin{eqnarray}
r_*=\int \frac{r^2}{\Delta} dr, \end{eqnarray}
 and resolving Eqs. (\ref{T1}) and (\ref{T2}) in the form
 \begin{eqnarray}
&&R_{+1/2}=\frac{\Delta^{-1/4}}{r}(\lambda^2+\mu^2
r^2)^{1/4}e^{-i\vartheta/2} \Psi_+,\nonumber
\\ &&R_{-1/2}=\frac{\Delta^{1/4}}{r}(\lambda^2+\mu^2
r^2)^{1/4}e^{i\vartheta/2}\Psi_-,
 \end{eqnarray}
with
 \begin{eqnarray}
 \vartheta=\arctan(\mu r/\lambda),
  \end{eqnarray}
 we obtain two wave-equations
\begin{eqnarray}
&&  \frac{d^2 \Psi_+}{d r_*^2}+\left\{\frac{d H_+}{d
r_*}-H_+^2+\frac{\Delta}{r^4}P_+\right\}\Psi_+
 =0,\label{LV1}\\
 &&  \frac{d^2 \Psi_-}{d r_*^2}+\left\{\frac{d H_-}{d
r_*}-H_-^2+\frac{\Delta}{r^4}P_-\right\}\Psi_-
 =0,  \label{LV2}
 \end{eqnarray}
where
 \begin{eqnarray}
&&H_-=\frac{1}{4 r^2}\frac{d\Delta}{dr}-\frac{\Delta}{r^3}+
\frac{i\mu}{2(\lambda+i\mu r)}\frac{\Delta}{r^2}, \\
&&H_+=-\left[\frac{1}{4r^2}\frac{d\Delta}{dr}+\frac{\Delta}{r^3}
+\frac{i\mu}{2(\lambda-i\mu r)}\frac{\Delta}{r^2}\right].
 \end{eqnarray}
Near the event horizon the asymptotic solutions are
\begin{eqnarray}
R_{\pm 1/2}\simeq e^{i\omega r_*}e^{\mp
\frac{i}{2}tan^{-1}\left(\frac{\mu r}{\lambda}\right)},~~~~
or~~~~R_{\pm 1/2}\simeq \Delta^{-s} e^{-i\omega r_*}e^{\pm
\frac{i}{2}tan^{-1}\left(\frac{\mu r}{\lambda}\right)}.
\end{eqnarray}
One may expand the wave-equations (\ref{LV1}) and  (\ref{LV2})  as
a power series in $M/r$ (neglecting terms of order
$O\left(\left(M/r\right)^2\right)$) as follows
\begin{eqnarray}
\label{VV1}&&\left[\frac{d^{2}}{dr^{2}}+\omega^{2}-\mu^2+
\frac{4M\omega^{2}-2M\mu^2}{r}-\frac{\lambda^2+\frac{\lambda}{\mu}\omega
}{r^{2}} \right] \xi_\pm=0,
\end{eqnarray}
where $\xi_{\pm}=\left(1-2 M/r\right)^{1/2}\Psi_{\pm}$. It is of
interesting to note that Eq. (\ref{VV1}) becomes Eq. (\ref{m19})
if we replace $\lambda$ by $-k$, which shows that the conclusions
of this paper do not change if we use usual tortoise coordinate
instead of the coordinate (\ref{tor}).

%\newpage

\newpage

\end{document}